\documentclass[prb,reprint,superscriptaddress,nobalancelastpage,twocolumn,showpacs]{revtex4-1}
\usepackage{color,amsthm,amsmath,amsfonts,graphicx,bm,amssymb,hyperref}
\usepackage{epstopdf}
\usepackage{comment}
\usepackage{etoolbox}
\begin{document}

\preprint{APS/123-QED}

\title{Stationary states in a free fermionic chain from the Quench Action Method}

 \author{Andrea De Luca}%
 \email{andrea.de.luca@lpt.ens.fr}
 \affiliation{%
 Laboratoire de Physique Theorique de l'ENS \& Institut Philippe Meyer,  Paris  - France 
}%
\author{Gabriele Martelloni}
\email{gabriele.martelloni@gmail.com}
\thanks{\textit{Present address:} SISSA, Trieste, Italy. }
\affiliation{%
Dipartimento di Fisica dell'Universit\`a di Pisa and INFN, Pisa - Italy 
}%
 \author{Jacopo Viti}
 \email{jacopo.viti@lpt.ens.fr}
\thanks{\textit{Present address:} Max Planck Institut for Complex Systems, Dresden, Germany. }
 \affiliation{%
 Laboratoire de Physique Theorique de l'ENS \& CNRS, Paris - France 
}%



\date{\today}

\begin{abstract}
We employ the Quench Action Method (QAM) for a recently considered geometrical quantum quench: 
two free fermionic chains initially at different temperatures are joined together in the middle and let evolve unitarily with a translation
invariant Hamiltonian. We show that two different stationary regimes are reached at long times, depending on the interplay between
the observation time scale $T$ and the total length $L$ of the system. We show the emergence of a non-equilibrium steady state (NESS) supporting an energy current for
observation time $T$ much smaller than the system size $L$. We then identify a longer time-scale for which thermalization occurs in a Generalized Gibbs Ensemble (GGE).

\end{abstract}

\maketitle

\newcommand{\beq}{\begin{equation}}
\newcommand{\eeq}{\end{equation}}
\newcommand{\Ord}[1]{{O}\left(#1\right)}

\newcommand{\sign}{\,{\rm sign}}
\newcommand{\arctanh}{\,{\rm arctanh}}

\def\bra#1{\mathinner{\langle{#1}|}} 
\def\ket#1{\mathinner{|{#1}\rangle}} 
\newcommand{\braket}[2]{\langle #1|#2\rangle} 
\def\Bra#1{\left<#1\right|} 
\def\Ket#1{\left|#1\right>}
\providecommand{\abs}[1]{\lvert#1\rvert}  
\providecommand{\norm}[1]{\lVert#1\rVert}

\newcommand{\twovec}[2]{\left[\begin{array}{c}
#1\\#2
\end{array}
\right]}

\newcommand{\avg}[2]{\langle #1 \rangle_{\mbox{\tiny #2}}}
\def\Tr{\operatorname{Tr}}
\def\diag{\operatorname{diag}}

\def\frl{\phi_{r/l}}
\def\fr{\phi_{r}}
\def\fl{\phi_{l}}
\def\fRL{\psi_{R/L}}
\def\fR{\psi_{R}}
\def\fL{\psi_{L}}

\def\irl{\Phi_{r/l}}
\def\ir{\Phi_{r}}
\def\il{\Phi_{l}}
\def\iRL{\Psi_{R/L}}
\def\iLR{\Psi_{L/R}}
\def\tiRL{\tilde\Psi_{R/L}}
\def\iR{\Psi_{R}}
\def\iL{\Psi_{L}}
\newcommand{\ialpha}[1]{\Psi_{#1}}

\def\fLW{\omega}
\def\fFO{\Psi}
\def\fFW{\psi}
\def\fFOstat{\vec{\fFO}^{\mbox{\tiny stat}}}
\def\FRL{\vec{\Upsilon}}
\def\ttheta{\tilde\theta}

\def\rightP{\mathcal{P}}
\def\mysmall{\delta}

\def\MM{M}
\def\bigM{\mathcal{M}}
\def\bigN{\mathcal{N}}
\def\bigc{\mathbf{c}}
\def\bigQ{\mathcal{Q}}
\def\bigpsi{\mathbf{\Psi}}
\def\bigphi{\mathbf{v}}
\def\bigA{\mathcal{A}}
\def\bigB{\mathcal{B}}
\def\bigAt{\tilde{\mathcal{A}}}
\def\smallM{m}
\def\Hleft{H_l}
\def\Hright{H_r}
\def\smallh{h}
\def\betaleft{\beta_l}
\def\betaright{\beta_r}
\newcommand\pv[1]{P_v\left[#1\right]}
\def\mom{p}
\def\Li{\operatorname{Li}}
\def\rhostat{\hat\rho_{\text{\tiny stat}}}
\def\citare{{\color{red}[??]}}
\def\hH{\hat{H}}
\def\hO{\hat{\mathcal{O}}}
\def\Trev{T_{\text{rev}}} 
\def\Tth{T_{\text{th}}} 
\graphicspath{ {../} }
\def\partA{\mu} 
\def\partB{\nu} 

\newcommand\makered[1]{{\color{red} #1}}
\newcommand\makeblue[1]{{\color{blue} #1}}
\newcommand\repl[2]{\sout{#1}\makered{#2}}
\newcommand\mrepl[2]{\text{\sout{\ensuremath{#1}}}\makered{#2}}
\def\vmax{v_{\mbox{\tiny max}}}

\section{Introduction}
Quantum quenches are nowadays the paradigm to study the long-time dynamics of isolated many-body quantum systems.
Their special interest relies in the possibility of an analytical treatment, especially in low dimensions~\cite{Silva_RMP}
 and in their potential realization as experimental protocols with ultracold atoms~\cite{bloch2008many, kinoshita2006quantum,  *gring2012relaxation}. The real-time
simulation of many-body quantum systems is also a challenge for sophisticated numerical algorithms~\cite{Schollwoeck2011}. 
The main physical question  regards the possibility of an effective thermodynamic description of
a quantum system initially prepared  with the density matrix $\hat{\rho}_0$,
after unitary evolution with an Hamiltonian $\hat{H}$. 
Formally the long-time density matrix $\rhostat$   is defined from 
$\lim_{t\rightarrow\infty}\Tr[\hat{\rho}_0\hat{\mathcal{O}}(t)]\equiv\Tr[\rhostat\hat{\mathcal{O}}]$ and this limit
is believed to exist for any
local operator $\hat{\mathcal{O}}$. However the characterization of $\hat{\rho}_{stat}$
in terms
of few macroscopic parameters of the system and the role of integrability are still open problems~\cite{rigol2008thermalization, calabrese2011quantum,
*mussardo2013infinite, pozsgay2014correlations}. 
In a recent work~\cite{caux2013time}, an unified approach, dubbed the Quantum Action Method (QAM), has been proposed 
to theoretically address
the characterization of the steady density matrix after an integrable quench. This method
exploits the special properties of the thermodynamic
limit $N\rightarrow\infty$, $L\rightarrow\infty$ at fixed density $N/L$,
with $N$ and $L$  the
particle numbers and the volume of the system respectively.

Known examples of applications are
restricted to $1+1$ dimensions when $\hat{H}$ is Bethe Ansatz solvable~\cite{gaudin1983fonction} and we will also stick to this situation in this paper. 
Roughly speaking, in the thermodynamic limit the many-body Hilbert space trace, defining
the expectation value of  local observables, can be replaced  by
a functional integral over a set of smooth macroscopic densities  $\rho_n(\phi)$ of single-particle momenta,
weighted by an exponential factor $\exp\bigl(-N\mathcal{S_Q}[\rho_n]\bigr)$. The
functional $\mathcal{S_{Q}}[\rho_{n}]$ is called the Quench Action
and is stationary at large times when all the  $\rho_{n}(\phi)$'s  equal  $\rho_{S}(\phi)$.
The function $\rho_S(\phi)$  identifies a representative pure state
that reproduces all the macroscopic features of the mixed state $\hat{\rho}_{stat}$  reached by the system for large times. 
The QAM 
requires the knowledge of the thermodynamic limit of the overlaps between the
many-body initial state and the eigenstates of the Hamiltonian $\hat{H}$, a non-trivial task even in one-dimensional systems,
which has been solved in few particular cases~\cite{pozsgay2014correlations, brockmann2014gaudin, *brockmann2014quench, *bertini2014quantum}.

Here, we discuss an application of the QAM to a recently
considered~\cite{de2013nonequilibrium, collura2014quantum, de2014energy, karrasch2012nonequilibrium, castro2014thermodynamic,eisler2014area, *eisler2014entanglement} geometrical quench: two identical quantum spin chains
initially held at different inverse temperatures $\beta_{l/r}$ are joined together restoring translational invariance
and evolved unitarily. Extensions to higher dimensions are also of current interest~\cite{bhaseen2013far, *doyon2014non, collura2014non}.

In order to illustrate how the QAM can
be successfully employed to extract the stationary behavior of this system we focus on the simplest possible example:
a free fermionic  chain. 
Formally the steady state  displays different regimes  according to the order in which the large times and the thermodynamic
limit are taken. 
If the linear size of the system $L$ is sent to infinity before the observation time $T$,
 $\rhostat$ supports a persistent energy current~\cite{aschbacher2003non} . Concretely, there will be a range
 of times $T \ll L/\vmax $, where $\vmax$ represents the speed of the fastest mode of the system \cite{lieb1972finite, bonnes2014light}, for which
this Non-Equilibrium Steady State (NESS) is observable. For much longer times, boundaries start
to be relevant in the dynamics and  a complete time-reversal symmetric state is restored, where all the expectation values of
local operators coincide with the mean between their thermal average in the two disconnected chains. This regime can be observed  for $T\ll \Trev$,
where $\Trev \propto L^2$ is the typical revival time of a free fermionic chain~\cite{kaminishi2013recurrence, collura2014quantum}. 
As expected, thermalization occurs in a Generalized Gibbs Ensemble
(GGE). As we will show, both the two regimes are
captured by the QAM. 

The paper is organized as follow. In Sec. \ref{sec_model} we introduce our model:  the spin-$1/2$
XX-chain equivalent to a free-fermionic model after the Jordan-Wigner transformation. 
The overlaps between the Hamiltonian eigenstates before
and after the quench are shown to be determinants and need to be properly regularized
when taking the infinite volume limit $L\rightarrow\infty$.
In Sec. \ref{sec_Action} we carry out exactly the computation of the Quench Action and show the existence of 
two possible regimes for the stationary
state in Sec. \ref{sec_NESS} and \ref{sec_GGE}. Eventually, we summarize our findings in Sec. \ref{sec_conc}.

\section{The model}
\label{sec_model}
We consider two disconnected spin-$1/2$ XX chains with Hamiltonian $\hH_0=\hH_l+\hH_r$
\begin{align}
\label{XX}
& \hat{H}_{r}=\frac{1}{2}\sum_{n=1}^{L}(\hat\sigma_n^{x}\hat\sigma_{n+1}^{x}+\hat\sigma_n^{y}\hat\sigma_{n+1}^{y}),\\ 
& \hat{H}_{l}=\frac{1}{2}\sum_{n=1}^{L-1}(\hat\sigma_{-n}^{x}\hat\sigma_{-n+1}^{x}+\hat\sigma_{-n}^{y}\hat\sigma_{-n+1}^{y});
\end{align}
where $\hat\sigma_n^{\alpha}$ is a Pauli matrix at position $n\in\mathbb Z$ and $\alpha=x,y,x$.
The model is  equivalent to a
free fermionic chain exploiting the Jordan-Wigner transformation~\cite{lieb1961two}
\beq
\label{H_disc}
\hat{H}_{0}=-\sum_{n=1}^{L-1}(\hat c_{n}^{\dagger}\hat c_{n+1}+ \text{h.c.})-
\sum_{n=1}^{L-1}(\hat c_{-n}^{\dagger}\hat c_{-n+1}+\text{h.c});
\eeq
the operators $\hat c_n$ satisfy canonical anticommutation relations $\{\hat c_n,\hat c_m^{\dagger}\}=\delta_{nm}$. 
For $L\rightarrow\infty$, the quadratic form \eqref{H_disc} can be easily diagonalized introducing
single-particle fermionic operators in
momentum space through $\hat c_n=\sqrt{\frac{2}{\pi}}\int_{0}^{\pi}d\theta~\sin[\theta(n-1/2)] \hat\psi_{r}(\theta) $ for positive values
of $n$
and $\hat c_n=\sqrt{\frac{2}{\pi}}\int_{0}^{\pi}d\theta \sin[\theta(n-1/2)] \hat \psi_{l}(\theta)$ for negative. One finds
\begin{equation}
 \hH_{0}=\sum_{\lambda=l,r}\int_{0}^{\pi}d\theta~\varepsilon(\theta) \hat\psi_{\lambda}^{\dagger}(\theta)\hat\psi_{\lambda}(\theta),
\end{equation}
with $\varepsilon(\theta)=-2\cos\theta$ and the fermionic fields obeying
\beq
\{\hat\psi^{\dagger}_{l/r}(\theta),\hat \psi_{l/r}(\theta')\}=\delta(\theta-\theta').
\eeq
When the two chains are joined together and the infinite volume limit is considered,
the resulting Hamiltonian $\hH=\hH_0+\hat c^{\dagger}_0 \hat c_1+\hat c_1^{\dagger}\hat c_0$  is translation 
invariant and can be diagonalized by Fourier transform defining $\hat c_{n}=\int_{-\pi}^{\pi}\frac{d\phi}{\sqrt{2\pi}} e^{in\phi}\hat\psi(\phi)$. It follows
\beq
\label{Hfull}
\hH=\int_{-\pi}^{\pi} d\theta~\varepsilon(\phi) \hat \psi^{\dagger}(\phi)\hat \psi(\phi),
\eeq
with canonically normalized fields
\beq
\label{normalization}
\{\hat \psi^{\dagger}(\phi), \hat \psi(\phi')\}=\delta(\phi-\phi').
\eeq 
From the explicit expressions of the local fermions $\hat c_n$, we can 
derive the matrix elements
$M_{r/l}(\phi,\theta)\equiv\langle 0|\hat{\psi}(\phi)\hat{\psi}^{\dagger}_{r/l}(\theta)|0\rangle$ between the single-particle  fermionic
operators before and after the quench; $|0\rangle$ denotes the Fock vacuum. Their explicit form is as follows
\begin{align}
\label{ovr}
&M_r(\phi,\theta)=
\frac{1}{2\pi i}\left[\frac{e^{-i\theta/2}}{1-e^{i(\phi-\theta-i\delta)}}-\frac{e^{i\theta/2}}{1-e^{i(\phi+\theta-i\delta)}}\right],\\
\label{ovl}
&M_l(\phi,\theta)=\frac{1}{2\pi i}\left[\frac{e^{i\theta/2}}{1-e^{i(\phi+\theta+i\delta)}}-\frac{e^{-i\theta/2}}{1-e^{i(\phi-\theta+i\delta)}}
\right],
\end{align}
where $\delta>0$ is a small positive quantity needed to ensure convergence; physically it can be interpreted as an
infrared cutoff and we will see that the behavior of $\delta$ at large $T$ corresponds to different order of limits in $T$ and $L$.
Using Cauchy theorem  it is possible to show that in the limit $\delta\rightarrow 0^+$, (\ref{ovr}, \ref{ovl}) are consistent with the canonical normalization \eqref{normalization}.
A basis for the $N$-particle sector
of the fermionic  Fock space associated to the Hamiltonian \eqref{H_disc} is
$|\{\theta,\lambda\}\rangle_{N}\equiv|\theta_1^{\lambda_1},\dots,\theta_N^{\lambda_N}\rangle = \hat\psi^{\dagger}_{\lambda_1}(\theta_1)\ldots\hat\psi^{\dagger}_{\lambda_N}(\theta_N)|0\rangle$,
with $\lambda_i=l,r$. Similarly a basis for the $N$-particle
Fock space of the fermions after the quench is $|\Phi\rangle_{N}\equiv|\phi_1,\dots\phi_N\rangle = \hat\psi^{\dagger}(\phi_1)\ldots\hat\psi^{\dagger}(\phi_N)|0\rangle$. Then applying the Wick theorem to multipoint
correlation functions of the local fermions $\hat c_n$ one obtains
\beq
\label{det}
_N\langle\Phi|\{\theta,\lambda\}\rangle_M=\delta_{N,M}\det[M_{\lambda_{\partB}}(\phi_{\partA},\theta_{\partB})]_{\partA,\partB=1}^N.
\eeq
The overlap between the $N$-particle states before and after the quench is the determinant of an $N\times N$ matrix,
whose elements are given in (\ref{ovr}, \ref{ovl}).

\section{The Quench Action}
\label{sec_Action}
We now  derive the Quench Action for our protocol.
Initially the two halves are disconnected and independently thermalized,  the state of the system
is then described by the density matrix $\hat{\rho}_0=Z^{-1}e^{-\beta_l \hH_l}\otimes e^{-\beta_r \hH_r}$, with $Z$ a normalization constant.

Let us  consider the expectation value of a local operator $\hO$, evolved in time with the Hamiltonian \eqref{Hfull}.
Since the theory is
free  we can fix the particle number $N$ in the Fock spaces before and  after the quench and consider the limit $N\rightarrow\infty$ only at the very end. 
Such a treatment of the thermodynamic limit, in which the ratio $N/L$ is always zero,
fails to correctly treat interactions at finite particle density but is effective for free theories~\cite{leclair1999finite,*bullough1986quantum,*yang1969thermodynamics}. Formally we have
\beq
\label{formal_ev}
\Tr[\hO(t)\hat{\rho}_0]=
\sum_{\Phi,\Phi'}e^{-i(E_{\Phi}-E_{\Phi'})t}~_N\langle\Phi|\hO|\Phi'\rangle_N ~\mathcal{D}(\Phi,\Phi'),
\eeq
where the sum is over all the $N$-particle states $|\Phi\rangle_N$ with energy $E_{\Phi}=\sum_{i=1}^N\varepsilon(\phi_i)$ and we defined
\beq
\label{ov_average}
\mathcal{D}(\Phi,\Phi')=\sum_{\{\theta,\lambda\}}P\bigl(\{\theta,\lambda\}\bigr)~_N\langle\Phi|\{\theta,\lambda\}\rangle_N~_N\langle\{\theta,\lambda\}|\Phi'\rangle_N.
\eeq
The probability measure of the state $|\{\theta,\lambda\}\rangle_N$ is factorized
\beq
\label{prob}
P\bigl(\{\theta,\lambda\}\bigr)=\prod_{\partA=1}^{N}\frac{1}{\pi}[\delta_{\lambda_\partA,l}f_{l}(\theta_\partA)
+\delta_{\lambda_\partA,r}
f_{r}(\theta_\partA)]
\eeq
with $f_{l/r}(\theta)=1/(1+e^{\beta_{l/r}\varepsilon(\theta)})$, the usual Fermi-Dirac distribution at inverse temperature
$\beta_{l/r}$. The prefactor ensures the correct normalization since $\int_{0}^{\pi}d\theta f_{l/r}(\theta)=\pi/2$.

In order to compute \eqref{ov_average} we denote by $M_\partA(\theta_\partB|\lambda_\partB)$  the matrix
elements $M_{\lambda_\partB}(\phi_\partA,\theta_\partB)$ in \eqref{det} and observe that $\mathcal{D}(\Phi,\Phi')$ can be rewritten as the expectation
value with respect to the probability measure \eqref{prob} of the product of two determinants
\beq
\label{expect_value}
\mathcal{D}(\Phi,\Phi')=\mathbb E\bigl[\det[M_\partA(\theta_\partB|\lambda_\partB)]_{\partA,\partB=1}^N
\det[M^*_\partA(\theta_\partB|\lambda_\partB)]_{\partA,\partB=1}^N\bigr].
\eeq
The notation $\mathbb E[g]$ is a shorthand for
\beq
\mathbb E[g]\equiv\sum_{\lambda_1,\dots,\lambda_N}\int_{0}^{\pi}d\theta_1\dots\int_{0}^{\pi}d\theta_N~ P\bigl(\{\theta,\lambda\}\bigr)
g(\{\theta,\lambda\}\bigr).
\eeq
Expanding the two determinants in \eqref{expect_value} over permutations $\sigma, \omega\in S_N$ we get
\begin{align}
\mathcal{D}(\Phi,\Phi')&=\sum_{\sigma,\omega}(-1)^{\sigma+\omega} \mathbb E\left[\prod_{\partA=1}^N
M_{\sigma(\partA)}(\theta_{\partA}|\lambda_{\partA})M_{\omega(\partA)}^*(\theta_{\partA}|\lambda_\partA)\right]\nonumber\\
\label{Andreieff}
&=N!\det\left[I(\phi_\partA,\phi'_\partB)_{\partA,\partB=1}^N\right],
\end{align}
where the matrix elements $I(\phi,\phi')$ are now obtain expectation values in the single-particle space 
\beq
\label{integral_I}
I(\phi,\phi')=\sum_{\lambda=l,r}\int_{0}^{\pi}\frac{d\theta}{\pi}M_\lambda(\phi,\theta)M^{*}_\lambda(\phi',\theta)
f_{\lambda}(\theta).
\eeq
We note
in passing that \eqref{Andreieff} is known in Random Matrix Theory~\cite{Mehta2004} as \textit{Andr\'eief identity}~\cite{Andreief} and
requires independent matrix elements, e.g. free theories.

The integrals in \eqref{integral_I} can be computed exactly noticing that
$M_{r/l}(\phi,\theta)=-M_{r/l}(\phi,-\theta)$ and $f_{r/l}(\theta)=f_{r/l}(-\theta)$. Then the integration
domain can be extended to $\theta\in[0,2\pi]$ and becomes the unit circle $|z|=1$, in the complex variable
$z=e^{i\theta}$. Applying Cauchy theorem, taking care of the essential singularity of the functions $f_{r/l}(z)$
at $z=0$, one finds
\begin{multline}
I(\phi,\phi')=\frac{1}{4\pi^2}\left[\frac{f_r(\phi')-f_r(\phi)}{e^{i\phi'}-e^{-i\phi}}+
\frac{f_r(\phi')+f_{r}(\phi)}{e^{i(\phi'-\phi)+2\delta}-1}+\right.\\
\label{res_I}
\left.-\frac{f_l(\phi')-f_l(\phi)}{e^{i\phi'}-e^{-i\phi} }-\frac{f_l(\phi')+f_{l}(\phi)}{e^{i(\phi'-\phi)-2\delta}-1}\right].
\end{multline}
The matrix elements $I(\phi,\phi')$ have poles at $\phi=\phi'\pm 2i\delta$,
we remark that contrary to the single-particle overlaps (\ref{ovr}, \ref{ovl}) no
singularity is present for $\phi=-\phi'$.

The formal Fock-space trace \eqref{formal_ev} is in a free-theory a multidimensional integral over the
angular variables $\phi_\partA$ and $\phi_\partA'$, $\partA=1,\dots,N$, defining the momenta of the $N$-particle states
$|\Phi\rangle$ and $|\Phi'\rangle$. Let us denote those integrations shortly by $d^N\phi$ and $d^N\phi'$
and introduce a normalized momentum density  by
\begin{equation}
\label{density}
 \rho(\phi)=\frac{1}{N}\sum_{\partA=1}^{N}\delta(\phi-\phi_\partA),
\end{equation}
analogously we will also consider a function $\rho'(\phi')$. In the spirit of the QAM the double sum in
\eqref{formal_ev} is now replaced by a functional integral over the densities $\rho(\phi)$
and $\rho'(\phi')$
\begin{multline}
 \label{therm1}
 \Tr[\hat{\rho}_0 \mathcal{O}(t)]=
 N!\int{d}^N\phi\int d^{N}\phi'\int\mathcal{D}\rho~J[\rho]\int\mathcal{D}\rho'~J[\rho']~
 \\ e^{iNt\int d\phi(\rho-\rho')\varepsilon}
 ~_N\langle\Phi|\hat{\mathcal{O}}|\Phi'\rangle_N~\det\left[I(\phi_\partA,\phi'_\partB)\right]_{\partA,\partB=1}^N\;, 
\end{multline}
where $J[\rho]$ and $J[\rho']$  enforce the constraint \eqref{density}.
Following~\cite{dean2001extreme}, they can be replaced by a functional integral over auxiliary imaginary functions $g(\phi)$
and $g'(\phi)$ 
\begin{align}
\nonumber
J[\rho]&\equiv\delta(N\rho(\phi)-\sum_{i=\partA}^{N}\delta\bigl(\phi-\phi_\partA)\bigr)\\
\label{jacobian}
&=\int\mathcal{D}g~ e^{\int d\phi g(\phi)\bigl[N\rho(\phi)-\sum_{i=\partA}^{N}\delta(\phi-\phi_\partA)\bigr] }.
\end{align}
We then expand the determinant in \eqref{therm1} over permutations $\sigma\in S_N$ and observe that the permutation sign
$(-1)^\sigma$ can be absorbed inside the definition of the state $|\Phi'\rangle=(-1)^{\sigma}|\sigma\Phi'\rangle$,
with $|\sigma\Phi'\rangle\equiv|\phi'_{\sigma(1)}\dots\phi'_{\sigma(N)}\rangle$. The integration over $d^N\phi$ is now
factorized and one ends up with a term of the form
$\prod_{\partA=1}^N \int d\phi I(\phi,\phi_\partA)e^{-g(\phi)}$, that can
be exponentiated and rewritten in terms of the density $\rho'(\phi')$. Finally, similar manipulations
of the remaining integration over $d^N\phi'$ lead to
\beq
\label{pathintegral}
 \Tr[\hat{\rho}_0\hat{\mathcal{O}}(t)]=\int\mathcal{D}g\int\mathcal{D}g'\int\mathcal{D}\rho\int\mathcal{D}\rho'~h[\rho']
 e^{N\mathcal {S}_{\mathcal Q}[\rho,\rho',g,g']},
\eeq
where we absorbed a $(N!)^2$ prefactor as an inessential constant term inside $\mathcal{S}_{\mathcal{Q}}$. Notice that
since $\hat{\mathcal{O}}$ is a local operator 
the matrix element $_N\langle\Phi|\hat{\mathcal{O}}|\Phi'\rangle_N$ is a smooth function of the set of momenta
$\{\phi'\}$ that can be replaced for our purposes by a functional $h[\rho']$ appearing in \eqref{pathintegral}. 
The reason is that local observables can only affect microscopic details of the densities
$\rho', \rho$, which result in sub-leading contributions 
in the thermodynamic limit. A similar reasoning is standard in the study of  equilibrium properties for
one-dimensional integrable
spin chains~\cite{vsamaj2013introduction}.
The functional $\mathcal{S}_{\mathcal Q}$ is the Quench Action 
\begin{multline}
\label{action}
\mathcal {S}_{\mathcal Q}[\rho,\rho',g,g']= \int d\phi~(g\rho+g'\rho')+\log\int d\phi~e^{-g'}+ \\
it\int d\phi~\varepsilon(\rho-\rho')+\int d\phi~\rho'(\phi)\log\int d\phi'~I(\phi',\phi)e^{-g(\phi')},
\end{multline}
which is explicitly time-dependent. 

In the $N\rightarrow\infty$ limit the path-integral \eqref{pathintegral} is dominated
by its saddle points, which physically identify for large $t$ the stationary distribution of momenta $\rho_S(\phi)$.

\section{Non-Equilibrium Steady State}
\label{sec_NESS}
We now show how the NESS already found in~\cite{aschbacher2003non}  can be nicely re-derived imposing stationarity of
the Quench Action \eqref{action} and is therefore exact in the thermodynamic limit.
\begin{figure}[t!]
\begin{center}
\includegraphics[width=0.75\columnwidth]{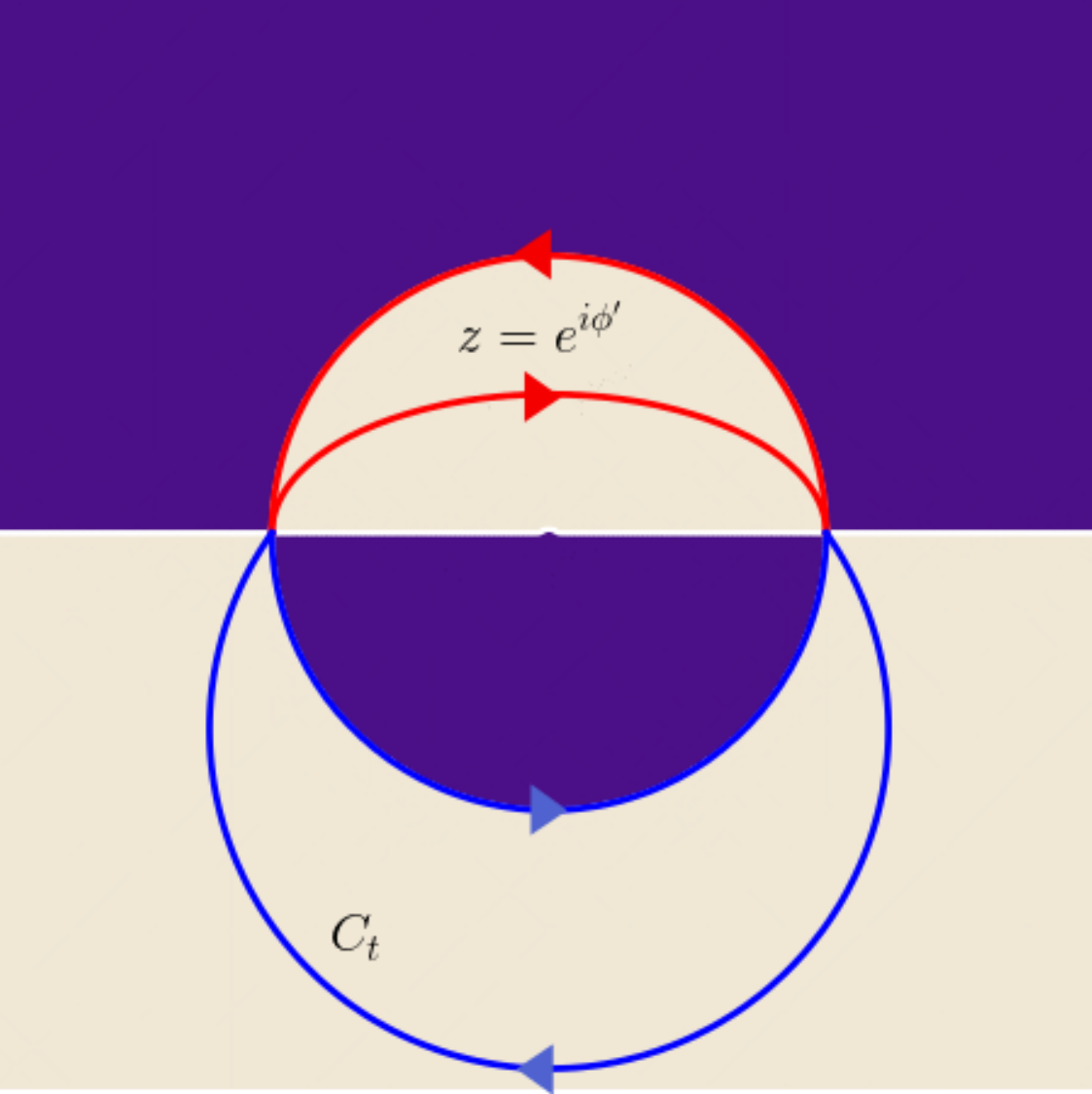}
\caption{The integration contour $C_t$ in the complex plane of $z=e^{i\phi'}$:
it consists of the blue and red closed curves depicted in figure.  The region where
$\Im[\varepsilon(z)]>0$ is colored in gray whereas the blue region corresponds to $\Im[\varepsilon(z)]<0$.
The contour $C_t$ avoids the
essential singularity at the origin of the integrand in \eqref{var_rhop}.} 
\label{fig_1}
\end{center}
\end{figure} 

Functional derivative of \eqref{action} with respect to $g'$ gives the condition
\beq
\label{minim_gp}
\left.\frac{\delta\mathcal{S}_{\mathcal{Q}}}{\delta g'}\right|_{\substack{\rho'=\rho'_S\\g'=g'_S}}=\rho'_S-\frac{e^{-g'_S}}{\int d\phi~ e^{-g'_S}}=0,
\eeq
implying the normalization of the stationary distribution $\rho'_S$ and reproducing the familiar entropic term
$-\rho'_S\log\rho'_S$ when substituted back in \eqref{action}.
The stationarity condition for $\rho$ simply identifies $g_S=-it\varepsilon$ and notice that consistently with \eqref{jacobian} this is a purely imaginary function.
Considering then the variation with respect to $\rho'$, with the aid of \eqref{minim_gp}, we arrive  at
\begin{multline}
\label{var_rhop}
\left.\frac{\delta\mathcal{S}_{\mathcal{Q}}}{\delta\rho'}\right|_{\rho'=\rho'_S}=
-\log\rho'_S(\phi)-it\varepsilon(\phi)\\ +\log\int d\phi' I(\phi',\phi) e^{it\varepsilon(\phi')}=0.
\end{multline}
The integral in \eqref{var_rhop} can be computed in the $t\rightarrow\infty$ limit extending it to the  complex plane of
$z=e^{i\phi'}$. For large times, line integrals in the region $\Im[\varepsilon(z)]>0$ are
exponentially vanishing and we can replace the original contour on the unit circle by the contour $C_t$, considered in Fig.~\ref{fig_1}. The final integration path consists of two closed curves:  one inside the unit circle (red-colored in Fig. \ref{fig_1})
for $\Im(z)>0$, the other outside the unit circle
(in blue in Fig.~\ref{fig_1}) when $\Im(z)<0$. Looking at the expression \eqref{res_I}, one realizes that
the pole inside the unit circle, with residue proportional to $f_r(\phi)$, contributes to the integral only of $\phi>0$;
vice-versa the pole outside the unit circle with residue proportional to $f_l(\phi)$ must be taken into account only
when $\phi<0$. Taking the limit $\delta\rightarrow 0^{+}$ we conclude from \eqref{var_rhop}
\beq
\label{ness}
\rho'_S(\phi)=\frac{1}{\pi}[\Theta(\phi)f_r(\phi)+\Theta(-\phi)f_l(\phi)],
\eeq
where $\Theta(\phi)$ is the Heaviside theta function.
Finally, the variation with respect to $g$ identifies $\rho'_S(\phi)$ with $\rho_S(\phi)$.

Any pure state defined by the momenta distribution in \eqref{ness} would be
a representative state for the factorized NESS encountered in free theories and CFTs~\cite{bernard2012energy, *bernard2013non}. Notice that the
derivation requires $\delta\rightarrow 0^+$ faster than $t\rightarrow\infty$, otherwise the contour integral along
$C_t$ would be vanishing. Physically this is related to the existence of the NESS for observation
times $T\ll L/\vmax$. It is not difficult to show that the density distribution \eqref{ness} implies
a finite and constant energy flow between the two halves of the chain~\cite{de2013nonequilibrium}.
\section{Thermalization time scale and GGE}
\label{sec_GGE}
A different reasoning is required to obtain the stationary state observed in~\cite{collura2014quantum}. 
In our protocol, where the initial state presents two subsystems that have a macroscopic energy difference,
thermalization requires a time $\Tth \ll \Trev$ that however diverges with the system size and in particular $\Tth \gg L/\vmax $. 
In this time regime, the previous approach is not correct since it assumed that the system was strictly speaking in the thermodynamic limit.
At finite $L$, the Hamiltonian \eqref{XX} can still be diagonalized in a similar manner.
The fermion momenta on the two halves are quantized according to $\theta^{(k)}=\frac{\pi k}{L/2+1}$,
with $k=1,\dots, L/2$ and have single-particle dispersion relation $\varepsilon(\theta^{(k)})=-2\cos\theta^{(k)}$; 
the Hamiltonians $H_{l/r}$ are then diagonal in the fermionic operators defined as 
\begin{align}
&\hat{\psi}_{r;k}=
\frac{2}{\sqrt{L+2}}\sum_{j=1}^{\frac{L}{2}}\sin(\theta^{(k)} j)~\hat{c}_j,\\
&\hat{\psi}_{l;k}=\frac{2}{\sqrt{L+2}}\sum_{j=0}^{-\frac{L}{2}+1}\sin\bigl[\theta^{(k)}(j+L/2)\bigr]~\hat{c}_j \;.
\end{align}
They satisfy canonical anticommutation relations $\{\hat{\psi}_{l/r;k},\hat{\psi}_{l/r;k'}^{\dagger}\}=\delta_{k,k'}$. 
For large $L$, the specific details of the boundary conditions are immaterial and for simplicity 
we choose periodic boundary conditions for the full chain obtained after joining the two halves 
in the middle. One gets the momenta $\phi^{(m)}=\frac{2\pi m}{L}$,
$m=-L/2+1,\dots, L/2$ with
dispersion relation $\varepsilon(\phi^{(m)})=-2\cos\phi^{(m)}$ and
the Hamiltonian becomes diagonal in the operators
\begin{equation}
\hat{\psi}_m=\frac{1}{\sqrt{L}}
\sum_{j=-\frac{L}{2}+1}^{\frac{L}{2}}e^{ij\phi_m}~\hat{c}_j,\quad \{\hat{\psi}_{m},\hat{\psi}_{m'}^{\dagger}\}=\delta_{m,m'}.
\quad. 
\end{equation}
We now reconsider \eqref{formal_ev}. Although at finite $L$, a similar 
reasoning can be used to write $\mathcal{D}(\Phi, \Phi') = N! \det[ I_L(\phi_\partA, \phi_\partB')]_{\partA,\partB=1}^N $.
Now $\Phi, \Phi'$ represents many-body configuration of particles with quantized momenta, while 
$I_L(\phi_\partA, \phi_\partB')$ replaces the result in \eqref{integral_I} with the finite $L$ thermal average
\begin{equation}
 \label{IfinL}
I_L(\phi,\phi')= \frac{2}{L}\sum_{\lambda=l,r}\sum_{k} f_{\lambda}(\theta^{(k)})M_\lambda(\theta^{(k)},\phi)M_\lambda^{*}(\theta^{(k)},\phi')
\end{equation}
where the prefactor comes from the normalization $\sum_k f_{l/r} (\theta_k) \simeq L/4$ and 
the finite $L$ overlaps take the form                                            
\begin{align}
\label{spo1}
&M_r(\theta^{(k)},\phi)=\frac{2}{\sqrt{L(L+2)}}\sum_{j=1}^{L/2}\sin(j \theta^{(k)})e^{ij \phi}\;,\\
\label{spo2}
&M_l(\theta^{(k)},\phi)=\frac{2}{\sqrt{L(L+2)}}\sum_{j=L/2 +1}^{L}\sin(j\theta^{(k)})e^{ij \phi} \;.
\end{align}
The derivation of the quench action can formally proceed along the same lines described in sec. \ref{sec_Action}. 
One arrives  to an equation analogous to \eqref{var_rhop}
\begin{equation}
\label{var_rhopdiscr}
\rho'_S(\phi)=\sum_{m} I_L(\phi^{(m)},\phi) e^{it(\varepsilon(\phi^{(m)})-\varepsilon(\phi))}.
\end{equation}
In the limit of large times, the sum is  dominated
by the value where the phase vanishes, i.e. $\rho'_S(\phi) \simeq \frac{L}{2\pi} I_L(\phi,\phi)$. 
The same conclusion follows taking the time average of \eqref{var_rhopdiscr}, 
and recalling that in the thermodynamic limit a Jacobian factor $\frac{L}{2\pi}$ is produced passing from the discontinuous function \eqref{density}
to the smooth density $\rho_S'(\phi)$.
In order to compute $I_L(\phi,\phi)$ for the large $L$, we notice that 
normalization ensures that 
\begin{equation}
\sum_k |M_{r/l}(\theta_k,\phi_m)|^2= 1/2 
\end{equation}
Moreover for large $L$ the support of $|M_{r/l}|^2$ concentrates in a window $|\theta^{(k)} - \phi^{(m)}| = O(L^{-1})$.
This two conditions are sufficient to see that 
\begin{equation}
 \label{rhoGGE}
 \rho'_S(\phi) = \frac{L}{2\pi} \lim_{L\to \infty} I_L(\phi, \phi) = \frac{1}{2\pi}(f_l(\phi) + f_r(\phi)) \;. 
 \end{equation}
This stationary state is clearly symmetric under parity $\phi \to -\phi$ and therefore no current is present. Moreover,
this result is perfectly consistent with the GGE prediction. Indeed, the local quench we studied does not affect
the expectation value of extensive conserved quantities, that therefore remain the sum of the contributions
of the two halves, i.e. $Q_{\text{\tiny tot}} = Q_l + Q_r$. For the free system we are considering, 
this requires that all the occupation number at the same energy 
from the two halves simply sum up, or equivalently \eqref{rhoGGE}.
\section{Conclusions}
In this paper we studied the behavior of two identical quantum XX chains, 
initially thermalized at different inverse temperatures $\beta_{l/r}$ and then suddenly put in contact in the middle. 
We formulated the problem of determining the stationary state reached by the system at large times after the quench in
terms of the recent QAM and showed that two different regimes are possible.
A genuine NESS, characterized by time reversal symmetry breaking and the emergence of 
a stationary energy current, describes the chain for observation times $T\ll L/\vmax$.  
At larger times the system is locally equivalent to a GGE:  a state that indicates thermalization
constrained to conservation laws where no current is flowing. 
These two scenarios are captured by the  QAM and our analytic results completely agree with the numerics in~\cite{collura2014quantum}. 
We believe that the derivation of the Quench Action for this problem is a promising and necessary step
in the attempt to generalize the calculation to interacting models; notably the Lieb Liniger gas.
In particular the latter is currently under investigation by the authors. 
\label{sec_conc}
\vspace{0.5cm}

\noindent
\textbf{Acknowledgements ---} We are grateful to  Pasquale Calabrese and Benjamin Doyon for interest in 
this work. GM and JV would like to
thank the ICTP of Trieste for hospitality. The second author
thanks Mario Collura for helpful discussions and acknowledges the ERC for financial support under Starting Grant 279391 EDEQS.

\bibliography{noneq_NESS}

\end{document}